\documentclass[
    english,
    reprint,
    longbibliography,
    notitlepage,
    superscriptaddress,
    preprintnumbers,
    prb,
    nobibnotes,
]{revtex4-2}

\usepackage{graphicx}
\usepackage{bm}
\usepackage{hyperref}
\usepackage{amsmath}
\usepackage{amssymb}
\usepackage{microtype}
\usepackage[dvipsnames, svgnames]{xcolor}
\usepackage[capitalise]{cleveref}
\usepackage{bbm}
\usepackage{mathtools}
\usepackage{dutchcal}
\usepackage{tensor}
\usepackage{newfloat}
\usepackage{subcaption}
\usepackage{soul}
\usepackage{ulem}
\usepackage{gensymb}
\usepackage{tipa}
\usepackage{ragged2e}
\usepackage{xspace}
\usepackage{enumitem}

\definecolor{lightgray}{gray}{0.9}
\definecolor{Amber}{rgb}{1.0, 0.75, 0.0}
\definecolor{blizzardblue}{rgb}{0.67, 0.9, 0.93}
\definecolor{burningsand}{RGB}{220, 148, 129}
\definecolor{burgundy}{rgb}{0.5, 0.0, 0.13}
\definecolor{darkgreen}{RGB}{6, 114, 43}

\DeclareFloatingEnvironment[
    fileext=lob,
    listname={List of Boxes},
    name=Box,
    placement=tbp,
]{floatbox}

\makeatletter
\def\caption@justification{%
    \rightskip\z@skip
    \leftskip\z@skip
    \parfillskip=0pt plus 1fil
    \relax
}
\makeatother

\DeclareTextFontCommand{\emph}{\itshape}

\makeatletter
\renewcommand{\fnum@figure}{Fig. \thefigure}
\makeatother

\newcommand*{\cmb}{\textsc{CMB}}
\newcommand*{\flrw}{\textsc{FLRW}}

\newcommand*{\lcdm}{\(\Lambda\)CDM}
\newcommand*{\lss}{\textsc{LSS}}
\newcommand*{\kl}{\textsc{KL}}
\newcommand*{\nasa}{\textsc{NASA}}

\newcommand*{\compact}{\textsc{COMPACT}}

\newcommand*{\cobe}{\textsc{COBE}}
\newcommand*{\wmap}{\textsc{WMAP}}
\newcommand*{\Planck}{\textit{Planck}\xspace}
\newcommand*{\litebird}{LiteBIRD}
\newcommand*{\taurus}{\textit{Taurus}}

\makeatletter
\renewcommand\paragraph[1]{%
    \par\addvspace{1ex}% space before (adjust if you like)
    \noindent\textbf{#1}\hspace{0.6em}\ignorespaces%
}
\makeatother

\let\orgautoref\autoref
\renewcommand{\autoref}{%
    \def\equationautorefname{Eq.}%
    \def\figureautorefname{Fig.}%
    \def\boxautorefname{Box}%
    \def\sectionautorefname{Section}%
    \def\subsectionautorefname{Section}%
    \def\subsubsectionautorefname{Section}%
    \orgautoref
}

\newcommand*{\E}[1]{\texorpdfstring{\ensuremath{E_{#1}}}{E#1}}
\newcommand{\slabi}{\texorpdfstring{\ensuremath{\E{16}^{(\mathrm{i})}}}{E16i}}% Inhomogeneous E16

\DeclareMathAlphabet{\mathpzc}{OT1}{pzc}{m}{it}

\makeatletter
\DeclareRobustCommand{\rcite}[1]{%
    \rcite@aux#1,\@nil{#1}%
}
\def\rcite@aux#1,#2\@nil#3{%
  \if\relax#2\relax
    Ref.~\cite{#3}%
  \else
    Refs.~\cite{#3}%
  \fi
}
\makeatother

\hypersetup{
    colorlinks=true,
    citecolor={blue},
    linkcolor={blue},
    urlcolor={blue},
}

\begin{document}

\title{The Topology of the Universe}

\author{Craig~J.~Copi}
\affiliation{CERCA/ISO, Department of Physics, Case Western Reserve University, 10900 Euclid Avenue, Cleveland, OH 44106, USA}

\author{Deyan~P.~Mihaylov}
\email{deyan.mihaylov@case.edu}
\affiliation{CERCA/ISO, Department of Physics, Case Western Reserve University, 10900 Euclid Avenue, Cleveland, OH 44106, USA}
\affiliation{Department of Astronomy, Faculty of Physics, Sofia~University~``St.~Kliment~Ohridski'',\\5 James Bourchier Blvd, 1164 Sofia, Bulgaria}

\author{Anna~Negro}
\affiliation{CERCA/ISO, Department of Physics, Case Western Reserve University, 10900 Euclid Avenue, Cleveland, OH 44106, USA}

\author{Amirhossein~Samandar}
\affiliation{CERCA/ISO, Department of Physics, Case Western Reserve University, 10900 Euclid Avenue, Cleveland, OH 44106, USA}

\author{Glenn~D.~Starkman}
\email{glenn.starkman@case.edu}
\affiliation{CERCA/ISO, Department of Physics, Case Western Reserve University, 10900 Euclid Avenue, Cleveland, OH 44106, USA}

\author{Yashar~Akrami}
\affiliation{Instituto de F\'isica Te\'orica (IFT) UAM-CSIC, C/ Nicol\'as Cabrera 13-15, Campus de Cantoblanco UAM, 28049 Madrid, Spain}
\affiliation{CERCA/ISO, Department of Physics, Case Western Reserve University, 10900 Euclid Avenue, Cleveland, OH 44106, USA}
\affiliation{Astrophysics Group \& Imperial Centre for Inference and Cosmology, Department of Physics, Imperial College London, Blackett Laboratory, Prince Consort Road, London SW7 2AZ, United Kingdom}

\author{George Alestas}
\affiliation{Instituto de F\'isica Te\'orica (IFT) UAM-CSIC, C/ Nicol\'as Cabrera 13-15, Campus de Cantoblanco UAM, 28049 Madrid, Spain}

\author{Stefano~Anselmi}
\affiliation{INFN, Sezione di Padova, via Marzolo 8, I-35131 Padova, Italy}
\affiliation{Dipartimento di Fisica e Astronomia ``G. Galilei", Universit\`a degli Studi di Padova, via Marzolo 8, I-35131 Padova, Italy}
\affiliation{Laboratoire Univers et Th\'eories, Observatoire de Paris, Universit\'e PSL, Universit\'e Paris Cit\'e, CNRS, F-92190 Meudon, France}

\author{Javier~Carr\'on~Duque}
\affiliation{Instituto de F\'isica Te\'orica (IFT) UAM-CSIC, C/ Nicol\'as Cabrera 13-15, Campus de Cantoblanco UAM, 28049 Madrid, Spain}

\author{Fernando~Cornet-Gomez}
\affiliation{Departamento de F\'isica, Universidad de C\'ordoba, Campus Universitario de Rabanales, Ctra. N-IV Km. 396, E-14071 C\'ordoba, Spain}

\author{Linn~Htat~Lu}
\affiliation{Astrophysics Group \& Imperial Centre for Inference and Cosmology, Department of Physics, Imperial College London, Blackett Laboratory, Prince Consort Road, London SW7 2AZ, United Kingdom}

\author{Andrew~H.~Jaffe}
\affiliation{Astrophysics Group \& Imperial Centre for Inference and Cosmology, Department of Physics, Imperial College London, Blackett Laboratory, Prince Consort Road, London SW7 2AZ, United Kingdom}

\author{Arthur~Kosowsky}
\affiliation{Department of Physics and Astronomy, University of Pittsburgh, Pittsburgh, PA 15260, USA}

\author{Mikel~Martin~Barandiaran}
\affiliation{Instituto de F\'isica Te\'orica (IFT) UAM-CSIC, C/ Nicol\'as Cabrera 13-15, Campus de Cantoblanco UAM, 28049 Madrid, Spain}
\affiliation{Departamento de F\'isica Te\'orica, Universidad Aut\'onoma de Madrid, 28049 Madrid, Spain}

\author{Thiago~S.~Pereira}
\affiliation{Departamento de F\'{i}sica, Universidade Estadual de Londrina, Rod. Celso Garcia Cid, Km 380, 86057-970, Londrina, Paran\'{a}, Brazil}

\author{Catherine~Petretti}
\affiliation{Harvard-Smithsonian Center for Astrophysics (CfA), 60 Garden St., Cambridge, MA 02138, USA}

\author{Andrius~Tamosiunas}
\affiliation{Institute of Theoretical Astrophysics, P.O. Box 1029 Blindern, N-0315 Oslo, Norway}

\collaboration{COMPACT Collaboration}

\preprint{IFT-UAM/CSIC-26-72 \quad\textbar\quad Invited Review for \textit{Nature Astronomy}}

\begin{abstract}\noindent
Is the Universe infinite in all directions? 
The only way to know is to look. 
A non-trivial cosmic topology would imprint subtle signatures on the cosmic microwave background (CMB) and on the three-dimensional distribution of matter, breaking statistical isotropy and, potentially, homogeneity at the largest scales. 
If the topology scale is small enough, these signatures would be observable. 
Over the past three decades, successive space missions, most notably WMAP and \textit{Planck}, have enabled sophisticated searches for these signatures, using methods ranging from looking for matched circle pairs to full Bayesian likelihood analysis based on topology-dependent covariance matrices. 
Although these searches have yielded no definitive evidence for non-trivial topology, current constraints exclude only some topologies, parameter ranges, and observer positions. 
Recent advances show that detectable signals may persist even when the topology scale exceeds the size of the visible Universe.
Planned CMB experiments, including LiteBIRD and \textit{Taurus}, and high-precision galaxy and line intensity-mapping surveys, could expand the detectable parameter space by exploiting polarisation data, and by exploring topology-induced correlations at all accessible redshifts.
Whether cosmic topology is observable remains uncertain, but current and future data offer an unprecedented opportunity to probe the global structure of the Universe.
\end{abstract}

%\keywords{cosmic topology, cosmic anomalies, cosmic microwave background, large-scale structure, statistical isotropy}

\begin{floatbox*}
\centering
\fbox{
    \begin{minipage}{0.95\textwidth}
        \justifying
        \noindent
        A \textbf{manifold} is a space which in a sufficiently small neighborhood of any point looks arbitrarily close to ordinary flat space. In the context of cosmic topology, spatial slices of the Universe are three-dimensional manifolds, while space-time is described as a four-dimensional manifold endowed with a Lorentzian metric. 
        (With a Lorentzian metric, the time contribution to the four-dimensional analogue of Pythagoras's theorem for the distance between two points enters with a negative sign.)
        We will focus exclusively on the properties of space, and leave aside all issues of time.
        \\[2pt]
        \noindent
        We refer to any of the three dimensions of space as a \textbf{compact dimension} if it has finite extent.
        A \textbf{compact manifold} has finite extent in every dimension. 
        \\[2pt]
        \noindent
        The (local) \textbf{geometry} of a manifold is specified by its metric structure and defines \textit{local} properties like curvature or how distances are measured. 
        We drop the word local for most purposes. \\[2pt]
        \noindent \(\bm{E^{3}}\), \(\bm{S^{3}}\), and \(\bm{H^{3}}\) denote the three maximally symmetric spatial geometries compatible with homogeneous and isotropic solutions of Einstein's equations. \(E^{3}\) is three-dimensional Euclidean (flat) space; \(S^{3}\) is the three-sphere, a three-dimensional space with constant positive curvature; and \(H^{3}\) is three-dimensional hyperbolic space, a three-dimensional space with constant negative curvature. \\[2pt]
        \noindent
        The \textbf{topology} of a manifold describes its \textit{global} characteristics, such as connectedness, compactness, and the existence of non-contractible loops. \\[2pt]
        \noindent
        The \textbf{covering space} is a simply connected space that locally reproduces the geometry of the Universe, from which the physical space is obtained by identifying equivalent points according to a discrete symmetry group. \\[2pt]
        \noindent
        The \textbf{fundamental domain} is a region of the covering space whose repeated copies, when appropriately identified, reproduce the entire physical Universe. \\[2pt]
        \noindent
        A \textbf{generator} is a transformation that identifies points of space while preserving distances and from which, together with the other generators, all the symmetry transformations of the space can be obtained by repeated application. \\[2pt]
        A \textbf{closed loop} is a continuous path on the spatial manifold whose initial and final points coincide. 
        Unshrinkable closed loops cannot be continuously deformed to a point and are a hallmark of non-trivial topologies. 
        \\[2pt]
        In a universe with unshrinkable closed loops, there is more than one straight path from an observer to an object at any other location. 
        We refer to the ``images'' of the object on all but the shortest path as \textbf{clones}.
    \end{minipage}
}
\caption{\textbf{Key definitions.}}\label{box:definitions}
\end{floatbox*}

\maketitle
\section{Why Cosmic Topology Matters}
\noindent
What is the topology of the Universe?

There is a long-standing misconception that the answer to that question can be obtained by measuring, for example using the cosmic microwave background (\cmb), the curvature of cosmic geometry over some suitably large volume and extrapolating. 
If the curvature is positive, then the Universe is the finite three-sphere \(S^{3}\), if the curvature is negative, then the Universe is the infinite hyperbolic three-space \(H^{3}\), and if the curvature vanishes, then the Universe is the infinite flat Euclidean three-space \(E^{3}\).
This is false. 
Topology is characterised by the possible existence and properties of unshrinkable closed loops---if you could travel far enough in some direction along the loop, you would return to your starting point. 
Different manifolds can have the same local geometry, i.e., the same metric, but different topologies. 
(See \autoref{box:definitions} for definitions of the key terms associated with spatial topology.)

In particular, this is true for manifolds with one of the three homogeneous and isotropic Friedmann-Lema\^{i}tre-Robertson-Walker (\flrw) spatial geometries of standard cosmology---\(S^{3}\), \(E^{3}\), and \(H^{3}\)\@. 
There exists a countable infinity of different topologies possible for manifolds admitting the \(S^{3}\) or \(H^{3}\) geometry. 
In contrast, there are only \(18\) topologies of Euclidean three-geometry, but the manifolds with those topologies are characterised by multiple real parameters.

Closed loops in \textit{spacetime} could arise because of very local short-distance exotic phenomena such as wormholes, which would be of great interest if discovered. 
However, it is far more conservative to imagine such loops being an intrinsic property of the \textit{spatial} geometry of the Universe on the very largest scales; we call this cosmic topology.

General relativity is a theory of the dynamical response of spacetime geometry and stress-energy to one another as encoded in the Einstein field equations.
These are a system of second-order non-linear coupled but local differential equations. 
Because they are local, these field equations give us little or no help in distinguishing among manifolds with different topology but the same local geometry. 
Cosmic topology is a global spatial property. 
To determine the cosmic topology we must, at the very least, measure a large enough region of the Universe such that its contents manifest effects of the boundary conditions topology imposes.

\paragraph{A long-standing question} 
The standard cosmological assumption is that the Universe contains no unshrinkable closed loops, meaning that we live in the covering space of one of \(E^{3}\), \(S^{3}\), or \(H^{3}\). 
Yet the history of questioning this assumption predates general relativity \cite{Schwarzschild:1900ueber, sommerville1914elements}. 
Soon after Einstein proposed the first fully relativistic model of cosmology, de Sitter criticised his choice of spatial manifold \cite{deSitter:1917zz}, suggesting that, instead of a simple three-sphere, one should consider a closely related topology in which opposite points on the three-sphere are identified. 
From a modern perspective, neither is clearly preferable to the other.
Interest in cosmic topology continued throughout the twentieth century: Friedmann recognised \cite{Friedmann:1924a} that \(E^{3}\) and \(H^{3}\) geometries admit compact manifolds, Lema\^{i}tre reiterated the point at the 1958 Solvay Conference \cite{1958_Lemaître_LaStructure}, and Heckmann and Sch\"{u}king highlighted it in L. Witten's classic \textit{Gravitation: An Introduction to Current Research} \cite{Witten1962gravitation}. 
Further discussions followed from Zeldovich, Sokoloff and Starobinsky \cite{Zeldovich} and from Ellis and Dyer at the 1987 Vatican Observatory Conference \cite{Stoeger1987}, while Fang and collaborators promoted cosmic topology throughout the 1980s and 1990s (e.g., Ref. \cite{Fang:1990cte}), as did H. Fagundes and his many colleagues in Brazil (e.g., Ref. \cite{Fagundes:1983cb}).

A popular article by the noted mathematicians J. Weeks and W. Thurston \cite{WeeksThurston1984} in the 1980s and a later comprehensive review by Lachi\`{e}ze-Rey and Luminet \cite{Lachieze-Rey:1995} helped inaugurate the modern era of sustained interest in cosmic topology. 
Considerable effort focused on cosmic crystallography, the search for repeated images of the same astrophysical objects across widely separated regions of the sky \cite{1987ApJ...322L...5F, Lehoucq:1996qe, Uzan:1999de}, a consequence of unshrinkable closed loops that allows light to reach the observer along multiple paths. 
These methods proved challenging \cite{Lehoucq:2000hf}, since multiple images would typically be seen at different cosmic times and from different viewing angles, and not all topologies produce signals in the pair-separation statistics once thought to be generic \cite{Gomero:1998dz}.

\paragraph{\textbf{Looking at the \cmb}}
After the first mapping of the \cmb\ by the Cosmic Background Explorer (\cobe) satellite in 1991, the potential of \cmb\ anisotropies to reveal the cosmic topology was quickly recognised, as was the advantage of having a source observed at the same cosmic epoch (and therefore at the same distance) in all directions.
While \cobe\ had detected primordial fluctuations in the \cmb\ intensity, the measurements remained limited by modest signal-to-noise ratio and restricted to relatively large scales, i.e., low harmonic number \(\ell\).
Despite these limitations, early attempts were made to use these low-\(\ell\) observations to place constraints on cosmic topology \cite{Starobinsky:1993yx, deOlivieraCosta:1994eb, deOliveira-Costa:1995vur, Bond:1996qs, Cornish:1997hz}.

It was soon recognised, however, that the upcoming Wilkinson Microwave Anisotropy Probe (\wmap), which was designed to produce a high signal-to-noise map of the \cmb\ over most of the sky up to \(\ell \lesssim 500\), would provide a much more robust data set for cosmic topology searches. 
In the years prior to \wmap's launch, numerous papers anticipated the scientific potential of its data, with much of these efforts devoted to generating simulated \cmb\ skies.
Given the strong priors at the time against a cosmological constant and the growing confidence that \(\Omega_{\mathrm{matter}} \simeq 0.3\), considerable attention was focused on the possibility that the Universe possessed \(H^{3}\) geometry. 
This possibility presented significant computational challenges for almost all non-trivial topologies. 
Those difficulties could be understood in a number of ways, but ultimately stem from the exponential growth of area and volume with distance in hyperbolic space. 
\wmap\ firmly established that the Universe is flat or nearly flat, which would alleviate many of those difficulties.

\begin{figure*}
    \includegraphics[width=\linewidth]{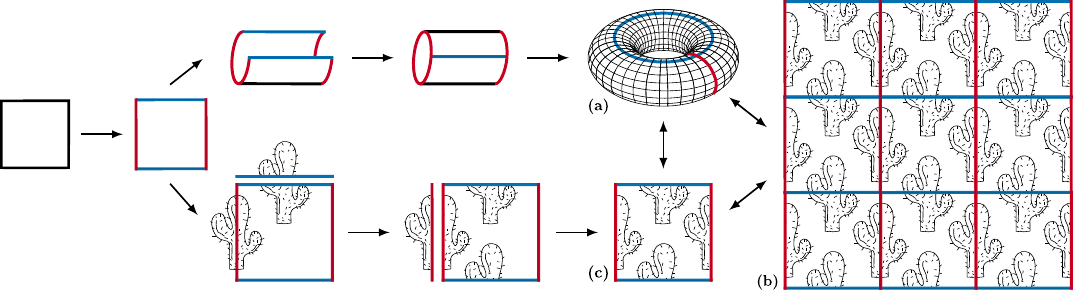}
    \caption{\textbf{Equivalent representations of a two-dimensional torus.}\hspace{0.6em} 
    A two-dimensional torus can be visualised in several ways: (a) starting from a square cell and gluing opposite sides to one another in a higher-dimensional flat space -- here three dimensions, for simplicity, but ideally in four dimensions to avoid the need to bend the surface; (b) tiling the plane with copies of the cell such that corresponding points in neighbouring copies are identified; and (c) imposing boundary conditions such that the glued sides correspond to “re-entering” the cell. 
    We note that the blue and red lines are identical in all three visualisations and are two representative unshrinkable closed loops, as is most clearly seen in (a).}
    \label{fig:visualizations}
\end{figure*}
\begin{figure*}[t]
    \centering
    \includegraphics[width=\linewidth]{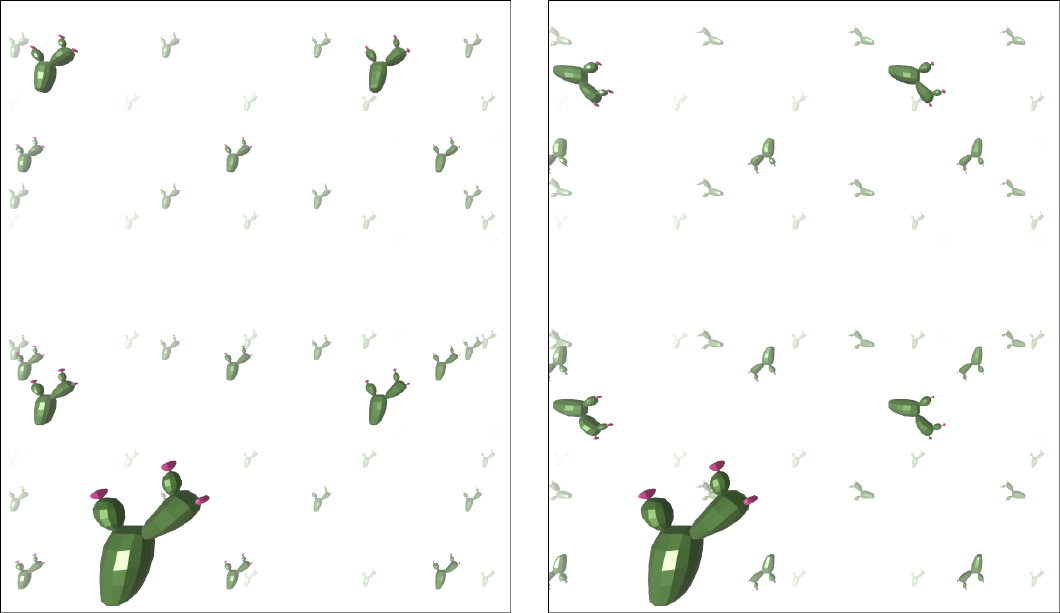}
    \caption{\textbf{A single cactus in a universe with non-trivial topology.}\hspace{0.6em} 
    We show the covering-space view with clones of the cactus related to one another by the group of transformations generated by three specific transformations, called generators.
    (An artificial fog is introduced to reduce the clutter caused by the infinite number of clones receding off to infinity.) 
    A fundamental domain would be any finite volume containing exactly one cactus and shaped so as to tile the covering space.
    \textit{Left panel:} The three generators of \E{1} are taken to be translations in the \(x\), \(y\), and \(z\) directions---i.e., upwards, rightwards, and into the page (shown, in perspective, as rightwards and \(\sim 30^{\circ}\) upwards).
    Note that with each step in any direction the cactus clone is unrotated. 
    The large cactus is nearby, as it is in the observer's fundamental domain, the other cacti are all clones, which appear to be more distant.
    \textit{Right panel:} The three generators of \E{3} are two orthogonal translations of equal length, here in the \(x\) and \(y\) directions (i.e., upwards and rightwards) and a corkscrew motion---a rotation by \(\pi/2\) followed by a translation in the \(z\) direction (shown, as in the left panel). 
    The translations are taken to be of the same length as in the left panel. 
    Note that with each step into the page the cactus clone is seen to rotate by \(\pi/2\). 
    For simplicity, the cactus has been placed on the axis of the rotation.\label{fig:topology}}
\end{figure*}

\section{Concepts in Cosmic Topology}

\noindent
There are several ways to visualize non-trivial topologies, three of which are illustrated in \autoref{fig:visualizations} for the two-dimensional torus. 
Each provides a different perspective on cosmic topology. 
We now discuss these visualisations in more detail, beginning with the observation that the Universe appears to be well described by one of the three \flrw\ geometries. 
For simplicity, we restrict our discussion to compact spaces, i.e., spaces with finite volume and finite spatial extent.

\begin{enumerate}[
    label=\textit{(\alph*)},
    leftmargin=17pt,
 ]
\item \textit{Identification in a higher dimensional space:} 
    If one imagines that there are more dimensions of space than the three that the Universe appears to have, then one can also imagine that the three-dimensional universe we perceive can be constructed by taking some finite piece of the full three-dimensional space \(E^3\), \(S^3\), or \(H^3\), and gluing the edges of that piece together to form a boundless shape that is our topologically non-trivial universe. 
    This is illustrated for a two-dimensional flat toroidal universe in \autoref{fig:visualizations} with the sequence of panels leading up to visualisation (a).

\item \textit{Tiling:}
    The covering space of the geometry can be tiled with appropriately shaped domains. For the two-dimensional torus this is represented in \autoref{fig:visualizations} (b). 
    These domains can be mapped into one another by elements of a discrete group of transformations that can be built out of (at least) 3 appropriately selected transformations called ``generators." 
    For example, in flat geometry (\(E^{3}\)) each of these generators involves an \(O(3)\) transformation (a rotation about an axis, reflection across a plane, or the identity) followed by a translation. 
    Select one such domain---designating it the \textit{fundamental domain}---and place ourselves (the observer) in it. Any physical event occurring inside this domain is replicated identically at the corresponding points of every other domain at precisely the same time.
    Topology thereby defines a universal notion of simultaneity \cite{Ellis1971}, in contrast to the local ``relativity of simultaneity'' at the core of special relativity \cite{poincare1898mesure, 1905AnP...322..891E} and 
    with interesting consequences for the famous twin paradox \cite{Uzan:2000wp, Barrow:2001rj}.) 
    Every physical field, represented by a function on spacetime, must be invariant under these transformations.
 
    We refer to copies of ourselves that appear in the other domains as \textit{clones}. 
    The fundamental domain itself has no physical significance---there are many differently shaped fundamental domains that describe exactly the same physical observations. 
    What matters physically are only the transformations that map one domain into another.
    \autoref{fig:topology} illustrates this conceptualisation of the simple three-torus and the ``quarter-turn space,'' 
    conventionally denoted \E{1} and \E{3} (not to be confused with the unrelated notation for the flat geometry \(E^{3}\)), respectively.
    For \E{1}, the 3 generators are 3 linearly independent translations, which, for simplicity, we have taken to be of equal length and in 3 orthogonal directions: \(x\) (upwards on the page), \(y\) (to the right on the page), and \(z\) (into the page).
    For \E{3}, we have taken the 3 generators to be a translation in the \(x\) direction, a translation in the \(y\) direction, and a translation in the \(z\) direction preceded by a \(\pi / 2\) rotation about the \(z\) axis (called a corkscrew motion). 
    We deliberately omit the nonphysical domain boundaries. 
    Instead, we show how repeated application of those generators moves a single object---a cactus---through the covering space. 
    (A cactus is chosen because it lacks symmetry under rotation or reflection.)
    The identical cactus is repeated ad infinitum. 
    Looking closely, we see that with each step to the right or upwards, the cactus is merely translated. 
    In \E{1}, the same is true for steps into the page, while in \E{3}, for each step into the page the cactus is rotated by \(\pi / 2\). (For simplicity, the cactus has been placed on the axis of that rotation.)
    An example of an unshrinkable closed loop would be any curve that connects one cactus to any other. 
    Representations of two unshrinkable closed loops on the two-dimensional torus are shown in \autoref{fig:visualizations} (a).

    While there is only a single cactus in each of these \E{1} and \E{3} universes, the images of cacti that appear further away correspond to light paths that traverse the universe one or more times before reaching us. 
    The apparent distance to an image corresponds to the actual distance the light has travelled, and different images are due to light leaving the cactus in different directions.
     
    Different choices of fundamental domains can be convenient for different purposes, provided the covering space can be tiled by application of the elements of the group of transformations. 
    Many etchings by M. C. Escher make use of this freedom to change the shape of the fundamental domain---replacing the simple parallelograms (in two dimensions) that a mathematician or physicist might draw with interlocking fish or birds.
    
\item \textit{Periodic domain:} 
    Consider any one of the fundamental domains of the tiling as a representation of the entire Universe (or, for a two-dimensional torus, \autoref{fig:visualizations}~(c)). 
    The allowed transformations then map the fundamental domain to itself, and map any point on the boundary of a fundamental domain to another point on the boundary of the same domain. 
    The closed loops we discussed above are then curves that start at a point, exit the domain through one boundary surface, re-enter through a different boundary, and eventually return to their starting point.
    If such a path can be continuously deformed into a spatial geodesic without being contractible to a point, then it is a closed loop.
    The familiar use of periodic boundary conditions on a cubic domain for many cosmological simulations is a very simple example of cosmic topology---the Euclidean 3-torus, a.k.a.\ \E{1} in~\autoref{fig:visualizations}.
\end{enumerate}

\paragraph{\textbf{Effects on the \cmb}}
In order to produce realisations of simulated \cmb\ skies, one must start by specifying the statistical properties of the fields that underlie the \cmb\ anisotropies. 
In standard cosmology, one key field is the primordial curvature potential. This is a scalar field typically expressed as a sum of spatial fluctuations that can exist in that geometry. In a standard inflationary universe, the amplitudes of those fluctuations are (very nearly) independent Gaussian random variables of zero mean with variances that depend only on their associated wavenumber. 
Because the underlying background geometry is homogeneous and isotropic, the resulting \cmb\ fluctuations are themselves statistically homogeneous and isotropic. 
For example, expanding the \cmb\ temperature fluctuations on the sky in spherical harmonics, their coefficients \(a^{T}_{\ell m}\) are independent Gaussian random variables whose ensemble average satisfies \(\left<a^T_{\ell m} a^{T*}_{\ell' m'}\right> = C^{TT}_{\ell} \delta_{\ell\ell'} \delta_{mm'}\).

The diagonality of the correlation matrix of \(a^{T}_{\ell m}\) in the covering space is a crucial calculational and computational simplification that relies on the statistical isotropy of the underlying geometry. 
A similar simplification in three dimensions---that the correlations between Fourier modes of the curvature perturbations are diagonal---relies on the homogeneity of the Universe. 
Cosmic topology, however, breaks isotropy. 
The fundamental domains are not rotationally invariant, the sole exception being de Sitter's preferred manifold. 
Cosmic topology typically also breaks homogeneity, violating the Cosmological Principle \cite{Fagundes:1991uy, Barrow:2003ma}.
For example, in \(E^{3}\) this occurs whenever any of the generators of the transformation group includes an \(O(3)\) element other than the identity, as they introduce a preferred axis of rotation or plane of reflection. 
These violations of isotropy and homogeneity create (typically small) correlations between previously 
independent modes;
thus \(\left<a^T_{\ell m} a^{T*}_{\ell' m'}\right> \equiv C^{TT}_{\ell m \ell' m'}\) is no longer a diagonal matrix---indeed, it is not even sparse.

\paragraph{\textbf{Interplay between geometry and topology}}
Pre-\wmap\ efforts focused on solving for the eigenmodes of the scalar Laplacian on manifolds with \(E^{3}\), \(H^{3}\), and \(S^{3}\) geometry and with non-trivial topology, then using these calculations to construct covariance matrices of \(a^T_{\ell m}\) (and also of \(E\)-mode polarisation coefficients \(a^{E}_{\ell m}\)), thereby enabling simulated realisations of the \cmb\ sky \cite{Riazuelo:2002ct}. 
This is an open-ended project. 
There are \(18\) distinct topologies of \(E^{3}\), conventionally labelled \E{1}--\E{18}. 
However, the eigenmodes, and thus the physics, requires more than simply identifying the topology---
one must also specify the translational vectors associated with each generator. 
This introduces up to 6 continuous parameters needed to describe the manifold, the eigenmodes, and the resulting correlation matrices of observables in a homogeneous \(E^{3}\) background geometry. 
Furthermore, observables depend on both the observer's orientation relative to these translations (up to 3 Euler angles) and the displacement of the observer from the axis of any rotation and the plane of any reflection (up to 3 more real parameters).

Topology on \(S^{3}\) is more challenging---there are five classes of such topologies \cite{Gausmann:2001aa}. 
Three have only one member, but two---the lens spaces and the prism spaces---each have countably infinite distinct members. Although the \(S^{3}\) manifolds of each topology have only one free real continuous parameter---the curvature scale---each topology yields different eigenvectors and correlation matrices of observables.

Topology on \(H^{3}\) is more challenging still \cite{Lachieze-Rey:1995, Levin:2001fg, thurston2022geometry}. 
For one, there are countably infinitely many topologies and they are unclassified; additionally, the number of clones grows exponentially with distance (because volume does in \(H^{3}\)), complicating both analytic and numerical calculations of eigenmodes and the statistical properties of observables. 

Significant effort from many different groups was put into exploring the curved non-trivial scenarios \cite{Inoue:1998nz, 1999math......6017C, Aurich:2000br, Uzan:2003ea, Aurich:2004xa, Caillerie:2007gd, Aurich:2012jc, Aurich:2012yd}, although the interest in these began to fade as more stringent constraints on the curvature of the Universe were inferred.

\paragraph{\textbf{Computing correlations}}
The correlation matrix of Gaussian random observables is no longer diagonal in universes with non-trivial topology.
Substantial progress has been made towards computing the correlation matrices of \cmb\ temperature harmonic coefficients arising from scalar fluctuations in the early Universe, both in the years preceding \wmap\ and subsequently, especially in preparation for the European Space Agency's \Planck\ satellite mission, but the task is still far from complete. 
Building on the work of others \cite{Riazuelo:2002ct, Riazuelo:2003ud, Lehoucq:2003math, Lehoucq:2002wy}, the \compact\ collaboration has set a goal of completing this work by generalising and extending previous computations \cite{COMPACT:2022gbl} and making the associated software publicly available in a GitHub repository \cite{compactgithub}.
For Euclidean manifolds, scalar eigenmodes and correlation matrices have now been computed in full generality \cite{COMPACT:2023rkp, COMPACT:2025adc}.
Also, for the first time, tensor eigenmodes and correlation matrices for all the orientable Euclidean manifolds (those in which the handedness of objects is not flipped by carrying them all the way around an unshrinkable closed loop) have been obtained \cite{Samandar:2025kuf, COMPACT:2024cud}, with the non-orientable Euclidean manifolds to be completed shortly
(A. Samandar \textit{et al.} (COMPACT) 2026, in preparation). 
The collaboration has also begun (M. Martin Barandiaran \textit{et al.} (COMPACT) 2026, in preparation) generalising earlier work on solving the scalar Laplacian eigenproblem for \(S^{3}\) manifolds including generalising the solutions to the tensor Laplacian \cite{Higuchi:1986wu}.
While methods to study eigenmodes of \(H^{3}\) manifolds and associated observables were developed \cite{Bond:1997ym, Bond:1999te, Bond:1999tf, Aurich:1999fh} and applied for select example manifolds prior to \wmap, these approaches remained comparatively slow.
There are plans to address the challenges of \(H^{3}\)---the much tighter constraints on spatial curvature today compared to the 1990s suggest that more efficient treatments may emerge from perturbing in the curvature.

\section{\cmb\ as a Probe of Cosmic Topology}
\begin{figure*}
    \includegraphics[scale=1]{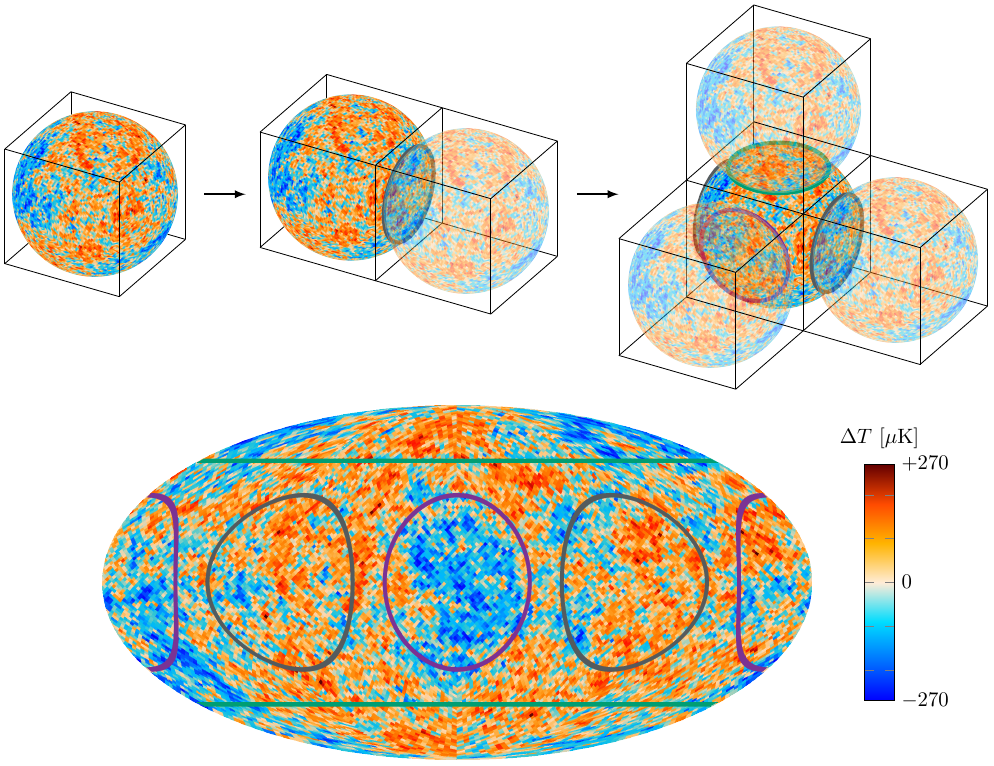}
    \caption{\textbf{Matched-circle pairs in the CMB}\hspace{0.6em} 
    Shown here as they would appear in CMB maps in the simple 3-torus topology \E{1}, with a cubic domain of dimension equal to 80\% of the diameter of the last scattering surface (LSS). 
    On the top left, the LSS is shown inside a cubic fundamental domain representing the periodic domain picture of cosmic topology as described in the text. 
    In the top centre, the overlap between the (spherical) LSSs of two nearest neighbours is shown, forming a circle that appears on both LSSs in different directions relative to the observer at the centre. 
    In the top right, we represent the tiled covering space picture of cosmic topology described in the text. 
    The matched-circle pairs on the LSS associated with nearest neighbour domains are shown. 
    (For clarity, only three of the six nearest neighbour domains are pictured but they are responsible for all three matched circle-pairs, all of which are depicted.) 
    At the bottom, the last scattering surface in a Mollweide projection is shown. 
    In this projection the matched-circle pairs are in the same colour: green for top and bottom, magenta for front and back, and grey for left and right.}
    \label{fig:circles}
\end{figure*}

\noindent
During the pre-\wmap\ era, it was recognised \cite{Cornish:1996kv, Cornish:1997ab, Cornish:1997rp} that, for isotropic metrics, there is a simple generic signature of cosmic topology that is present regardless of the specific topology.
Every observer sees the \cmb\ photons as having been emitted from a thin spherical shell centred on them---the last scattering surface (\lss) of \cmb\ photons. 
At any fixed cosmic time, the diameter of that \lss\ is independent of the location of the observer.
Consider the observer's nearest clone. 
Whenever the distance from the observer to that clone is less than the diameter of the \lss, the \lss\ of the observer and that of their clone intersect in a circle.
Both the observer and their clone see that circle, but in different directions on their respective skies, leading to a pair of circles observed in the sky. 
This is illustrated in \autoref{fig:circles} for a cubic 3-torus (\E{1}). 
However, since the observer and their clone are the same person making the same measurement, they are able to compare the patterns of temperature fluctuations around that pair of circles. 
If the only contribution to those fluctuations were the value of the curvature perturbation on the \lss, then the temperature fluctuations around those two circles would match exactly. 
(The relative sense of progression around the circles and the relative starting points would depend on details of the manifold.)
In practice, additional subdominant contributions---such as the velocity of the cosmic plasma at the time of last scattering, the changing gravitational potential along the line of sight to the \lss, absorption by dust, and gravitational lensing---reduce the correlation. Nevertheless, it was shown that, as long as the distance \(d_{\mathrm{NC}}\) to our nearest clone is less than roughly the diameter of the \lss, all such matched-circle pairs should, in principle, be detectable, enabling us to firstly detect non-trivial topology and secondly to reconstruct the topology of the Universe \cite{Weeks:1998qr}. 

\wmap\ was launched in 2001, and by the summer of 2002 a full-sky temperature map was available for the \wmap\ Science Team to analyse.
The search for matched-circle pairs was not straightforward.

Achieving optimal signal-to-noise ratio required locating the centres of the candidate circles to within less than \(1\degree\), specifying the circle radii with comparable accuracy, and determining the relative orientation of each circle pair to similar precision. 
Comparing each circle pair required pixelating each circle into segments of \(\leq 1\degree\). 
Independent implementation of the matched-circle search using \wmap\ (and later \Planck) data was carried out by several groups, exploring different statistical estimators \cite{Cornish:2003db, Roukema:2004iu, ShapiroKey:2006hm, Bielewicz:2010bh, Bielewicz:2012jnb, Vaudrevange:2012da}.
After an early indication of matched circles turned out to be a software error, the \wmap\ data were used to establish that the shortest distance around the Universe through us must exceed approximately 98.5\% of the diameter \(d_{\mathrm{LSS}}\) of the last scattering surface \cite{Vaudrevange:2012da}. 

Other potentially generic signatures of non-trivial topology (i.e., features that applied over a wide range of manifolds) were also commented on in pre-\wmap\ literature \cite{Levin:1997tu, Levin:1997uw, Levin:1998qu, Scannapieco:1998cz, Levin:1999bi, Olson:2000ft,Gomero:2001gq}; however, to the best of our knowledge, no dedicated searches exploiting these effects have been carried out using \cmb\ data.
Such suggestions continue to be made \cite{Aurich:2024bvu} and could eventually prove useful in the ongoing search for cosmic topology.

\wmap\ is rightly credited for ushering in the modern era of ``precision cosmology.'' 
Earlier balloon-borne and ground-based experiments had already improved on \cobe\ sufficiently to establish that the observable Universe is nearly flat (barring a Hubble constant far below the then-preferred value of \(\sim 70~\mathrm{km/s/Mpc}\)). 
\wmap, however, reduced uncertainties on the cosmological parameters of \lcdm\ cosmology to the \(1-10\%\) level. 
At the same time, \wmap\ revealed the existence of (or in some cases strengthened the case for) \textit{large-angle anomalies} \cite{Schwarz:2015cma, Planck:2013lks, Planck:2015igc, Planck:2019evm, Abdalla:2022yfr}---statistical evidence that the \cmb\ temperature was not the realisation of a Gaussian random statistically isotropic process. 
This evidence for statistical isotropy violation (some of it retrospectively identifiable in \cobe\ data) has remained robust ever since~\cite{Planck:2013lks, Planck:2015igc, Planck:2019evm}, contributing to the sustained interest in cosmic topology (although some cosmologists have advocated the possibilities that the anomalies are statistical flukes within \lcdm\ \cite{Pontzen:2010yw, Bennett:2010jb}). 
Several works \cite{Luminet:2003dx, Aurich:2021ofm} have attempted to draw a specific connection between topology and those large-angle anomalies, but so far always in contradiction to the conclusions of the \wmap\ matched-circle search.

\paragraph{\textbf{Cosmic topology in the \Planck\ era}}
The \Planck\ satellite was launched in 2009, approximately a decade after \wmap. 
\Planck\ improved on \wmap\ in several aspects potentially relevant to topology searches---higher signal-to-noise ratio across all topologically relevant angular scales, higher angular resolution, substantial sensitivity to polarisation, and more wavebands. 
For the matched-circle search in particular, the main advantage was the potential for increased confidence in the full-sky foreground-subtracted maps at multipoles near \(\ell \simeq 200\), especially with four independent foreground-subtraction methodologies 
each implemented by a different component-separation pipeline. 
The \Planck\ team repeated the matched-circle search, but only for circle pairs with antipodal centres \cite{Planck:2013okc, Planck:2015gmu}.
Neither this search nor an unpublished search over all possible circle pairs \cite{Vaudrevange}, carried out by the authors of \rcite{Vaudrevange:2012da}, found statistically significant evidence of matched circles.

Meanwhile, building on earlier work \cite{Phillips:2004nc, Niarchou:2007nn},
the \Planck\ team also conducted a likelihood-based search for cosmic topology \cite{Planck:2013okc, Planck:2015gmu}---they compared the likelihood that the observed part of the \cmb\ sky is a realisation of a topologically trivial universe versus that of a universe with non-trivial topology. Although they found no evidence for non-trivial topology, their analysis was limited to just two topologies with very restrictive parameter choices.
This method has a striking conceptual connection to Kac's famous mathematical question \cite{Kac:1966xd}: 
\textit{``Can one hear the shape of a drum?''}---that is, not with one's ears, but rather with perfect instrumentation. 
Mathematically, the answer is known to be ``no" \cite{Gordon:1992, 1996AmSci..84...46G}, since homophonic drums exist. 
From a physical perspective, however, most drum shapes---like the triangle, moon, and star illustrated---can be distinguished from one another. 
Since the \cmb\ anisotropies are the imprints of acoustic oscillations in the pre-recombination plasma, extracting the spatial topology from the \cmb\ is a close analogy, except that the drum is the whole three-dimensional Universe, the boundary conditions are topological rather than Dirichlet, and our instruments are imperfect. 
This is especially true if the last scattering surface lies entirely within a fundamental domain. 
In that case, any inference of topology must rely solely on the statistical properties of the amplitudes of the Laplacian eigenmodes. 
Existing work universally assumes that the standard inflationary prediction---that these are independent Gaussian random variables of zero mean---remains valid in topologically non-trivial cases, even though the presence of observable topology complicates the inflationary picture. In particular, it implies that inflation did not persist for a large number of e-folds of expansion, and therefore, is not, by itself, a complete explanation for the near-homogeneity of the Universe. 

\paragraph{\textbf{Quantum cosmology and the question of inflation}} At present, there is no established first-principles selection rule that singles out any particular spatial topology for the Universe. 
In quantum cosmology, proposals for the initial state, including the no-boundary, tunneling, and more recent boundary/end-of-the-world-brane constructions, provide different ways of assigning weights to possible cosmological manifolds \cite{Linde:2004nz, Hassfeld:2024hnx, Hartle:1983ai, Vilenkin:1982de, Maldacena:2024uhs}. 
More broadly, different quantum gravity approaches suggest that one may even extend the possible spaces beyond smooth manifolds with fixed topology \cite{Montero:2020icj, Hartle:1985nrz, Schleich:1993bs, Asselmeyer-Maluga:2010thh}. 
In summary, given the current understanding in a quantum gravity context, the Universe very plausibly starts out with a non-trivial topology that subsequent classical evolution does not alter.

In the very early Universe, when closed loops around the Universe are shortest,
topology will induce potentially observable quantum effects.
Non-trivial topology induces contributions to the stress-energy tensor \cite{Bire82, DeWitt:1979dd}, similar to the Casimir energy \cite{Casimir:1948dh} that arises most famously between parallel plate conductors.
While such effects have long been studied in quantum field theory in curved spacetime and in extra-dimensional compactifications, recent work has emphasised their potential role in early-Universe cosmology \cite{COMPACT:2026vdj,Fornal:2026qfq}. Establishing a quantitative connection between these quantum-topological effects and cosmological observables remains an area of active investigation, but it is expected that perturbations on scales comparable to the dimensions of the manifold will be altered from the standard inflationary predictions. 
The question of whether such effects are observable is again closely tied to the duration and initial conditions of inflation.
If inflation is brief enough for the topology itself to be observable, then the early-Universe quantum effects are likely to be detectable as well.

In an inflationary context, having a topological scale that is comparable to the Hubble scale appears to be a coincidence, requiring that the duration of inflation be neither too long---since the topological scale would then lie far beyond today’s Hubble radius---nor too short---since then it would be far smaller than the Hubble scale today. 
Instead it must be ``just right." 
This is the same as for spatial curvature, and, just as for spatial curvature, the apparent fine-tuning required for observable topology does not preclude its existence; rather, a detection would provide direct constraints on the duration of inflation and on pre-inflationary conditions. 
Again as for curvature, if there is observable topology, the pre-inflationary Universe must already have been highly homogeneous; otherwise the pre-inflationary deviations from the \flrw\ background would remain visible to us today. 
For negative, but not positive, curvature, there is a known possibility of the curved spacetime emerging quantum mechanically homogeneous and isotropic from a previous inflationary epoch \cite{Linde:1998gs}.
Similarly, topology offers a known potential mechanism for producing homogeneity through pre-inflationary mixing arising from the chaotic behaviour of geodesics on certain manifolds \cite{Misner:1969hg, Cornish:1996st, Barrow:1997sb, Barrow:1999sm}; this possibility awaits more comprehensive exploration. 
Alternately, the explanation for initial homogeneity may lie elsewhere, such as in quantum gravity.

If inflation both expanded the spatial manifold to its present observable size \textit{and} generated the primordial fluctuations, then we must revisit the predictions of inflationary models for topologically non-trivial space.

\paragraph{\textbf{Violation of statistical isotropy}} Adopting this inflation-inspired (though potentially more general) assumption that the amplitudes of scalar Laplacian eigenmodes are zero-mean independent Gaussian random variables whose variances depend only on the corresponding eigenvalues, it is possible to compute the expected correlation matrix of \(a^{T}_{\ell m}\) as a function of the topology of the spatial manifold and the parameters specifying both the manifold and the background cosmology. While in the simply connected covering space, this correlation matrix is diagonal, in topologically non-trivial manifolds, this diagonality is lost.

\begin{figure*}
    \includegraphics[scale=1]{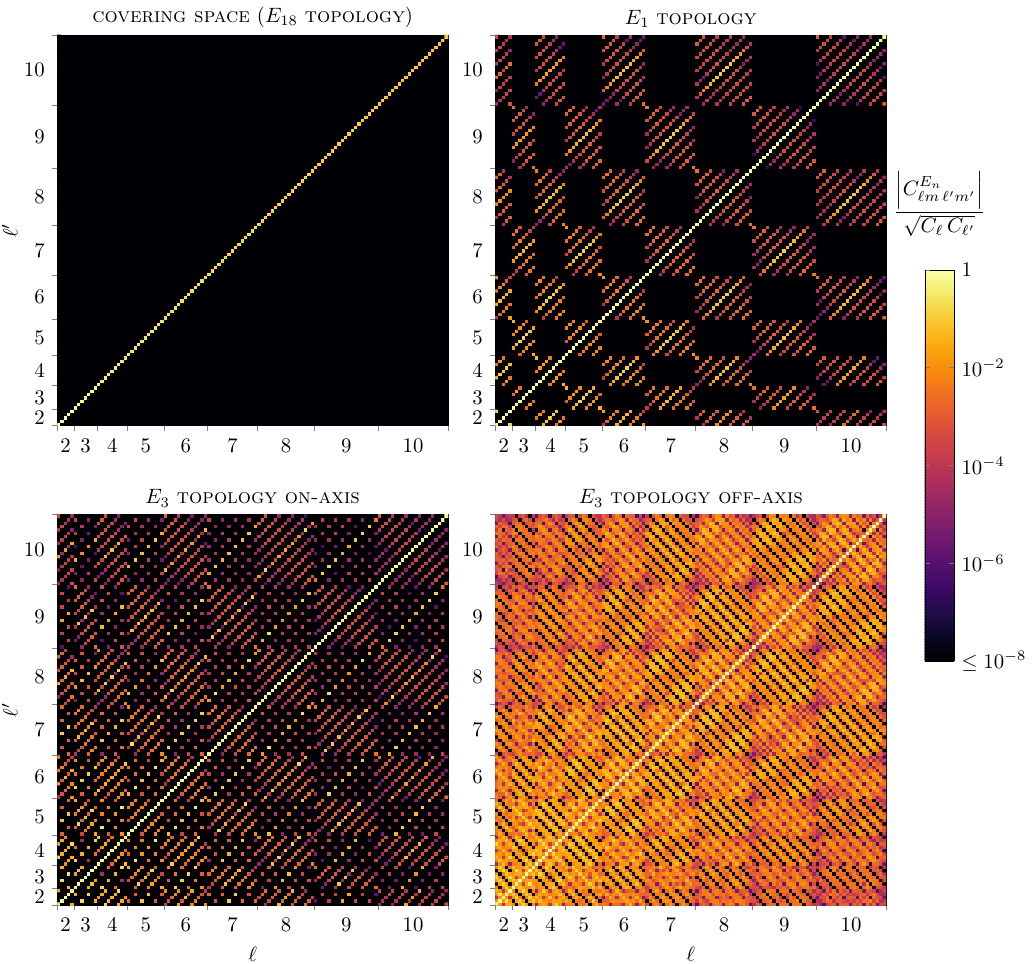}
    \caption{\textbf{\cmb\ temperature correlation matrices}\hspace{0.6em} 
    Absolute values of the rescaled \cmb\ temperature (\(TT\)) correlation matrices for topologies in which the length scale is chosen such that the diameter of the circles passing through the observer is 5\% larger than the diameter of the \lss: the covering space (also known as \E{18}, \textit{top left panel}), \E{1} topology (\textit{top right panel}); \E{3} topology with the observer on the axis of rotation (\textit{bottom left panel}); \E{3} topology with the observer off the axis of rotation (\textit{bottom right panel}).}
    \label{fig:corr_matrices}
\end{figure*}

The topological boundary conditions break isotropy, and so the correlation matrix \(C^{TT}_{\ell m\ell' m'}\) acquires diverse off-diagonal components, which depend on the specifics of each manifold and the position of the observer. 
\autoref{fig:corr_matrices} shows the resulting rescaled auto-correlation matrices of \cmb\ temperature fluctuations with \(2 \leq \ell \leq 10\) for four examples: the covering space \E{18} (\textit{top left panel}); the simple 3-torus \E{1} (\textit{top right panel}); the 3-torus with a \(\pi/2\) rotation \E{3} and the observer on the axis of that rotation (\textit{bottom left panel}); and finally the 3-torus with a \(\pi/2\) rotation and the observer off axis (\textit{bottom right panel}). 
Owing to its complete symmetry, \E{18} has the expected diagonal covariance matrix. \E{1} is homogeneous and preserves parity symmetry, so combinations with \(\ell + \ell'\) odd vanish. 
In contrast, the \E{3} topology breaks homogeneity and the correlation matrix \(C^{TT}_{\ell m\ell' m'}\) depends explicitly on the observer's position relative to the rotation axis. 
In all cases, any off-diagonal terms in the correlation matrix remain small compared to the diagonal ones, so a specific pattern can only be detected statistically over cosmic variance by averaging over many off-diagonal correlations. The likelihood statistic employs all the off-diagonal correlations within the \(\ell\) range to which it is applied.

Besides topology, other fundamental mechanisms are known to induce off-diagonal correlations between the coefficients \(a^T_{\ell m}\), such as anisotropic inflation \cite{Pitrou:2008gk}, non-Gaussian modulation \cite{Schmidt:2012ky}, and magnetic fields \cite{Durrer:1998ya}, to name a few. 
Unless finely tuned, however, it is unlikely that any combination of these mechanisms could produce correlations as rich as those seen in \autoref{fig:corr_matrices}. 
Systematic effects, such as the use of masks to reduce foreground contamination, also break statistical isotropy. 
Their effect, however, can be accurately modelled through simulations and the formalism of mode-coupling matrices \cite{Hivon:2001jp}.

By comparing the likelihood that the observed sky is a realisation of Gaussian fields with correlations determined by the computed topology-dependent covariance matrices, the \Planck\ team was unable to provide evidence that the observed \cmb\ is better explained by specific topologically non-trivial manifolds \citep{Planck:2013okc}. 
This was consistent with the negative circle searches, and, for certain topologies, allowed the limits on the topology scale to be extended to cases where the shortest distance around the Universe through us exceeded the diameter of the \lss.

These negative results dampened interest in cosmic topology, although at least part of that reaction was unwarranted. 
The \Planck\ Bayesian likelihood analysis yielded null results, however it was limited to a subset of Euclidean manifolds and only a restricted portion of their parameter spaces. 
\rcite{COMPACT:2023rkp, COMPACT:2025adc} undertook a more comprehensive exploration of the \(TT\) correlation matrices for Euclidean topologies. 
To assess the detectability of a certain topology, i.e., whether it could be distinguished from the covering space of \(E^{3}\), they computed the Kullback–Leibler (\kl) divergence \cite{kullback1951, kullback1959information} between the Gaussian distributions of \(a^{T}_{\ell m}\) defined by the correlation matrices, as done previously \cite{Kunz:2005wh, Fabre:2013wia} for a more limited set of topologies and manifold parameters.
They found that the \kl\ divergence can remain above unity (the conventional threshold for differentiating between two distributions) for manifolds---and observers within them---in which the shortest closed loop through the observer measures up to approximately \(1.3\) times the diameter of the \lss\ \cite{COMPACT:2023rkp, Samandar:2025kuf, COMPACT:2025adc}.

The results of the \Planck\ and \wmap\ circle searches did not yield positive results, but an important limitation of this method was often overlooked: it constrains only those closed loops that pass \textit{through our location}---a crucial observation since topology generically breaks statistical homogeneity. 
\rcite{COMPACT:2022nsu} examined the consequences of this fact for orientable Euclidean manifolds, showing that the length of the shortest closed loop around the Universe can be up to a factor of 6 smaller than the shortest closed loop through us. 
In one particular case, \(\slabi\), the rotated slab space, it could be arbitrarily shorter. 
In non-orientable Euclidean manifolds, the length of the shortest closed loop around the Universe can be up to a factor of 2 smaller than the shortest closed loop through us, see \cite{COMPACT:2025adc} and (Mihaylov, \textit{et al.} (COMPACT) 2026, in preparation).

\paragraph{\textbf{Renewed effort to search for cosmic topology}}
Members of the \compact\ collaboration are working to extend the \Planck\ likelihood-based topology search as far as computationally feasible.
The primary determinant of whether topology is detectable in cosmological data is the distance \(d_{\mathrm{NC}}\) to the observer's nearest clone compared to \(d_{\mathrm{LSS}}\). 
Matched circle searches of \wmap\ data were able to place limits \cite{ShapiroKey:2006hm, Vaudrevange:2012da} of \(d_{\mathrm{NC}} > 0.985 \, d_{\mathrm{LSS}}\), at least if that clone was not too close to the Galactic plane. 
Using a Bayesian likelihood search, the \Planck\ team was able, in particular topologies that they investigated \cite{Planck:2013lks, Planck:2015gmu}, to confirm this limit and even extend it as far as \(d_{\mathrm{NC}} \gtrsim 1.03 \, d_{\mathrm{LSS}}\). 
Preliminary estimates suggest that a comprehensive analysis of the \Planck\ data should be able to probe out to \(d_{\mathrm{NC}} \approx 1.05 \, d_{\mathrm{LSS}}\) for all manifolds with homogeneous and isotropic local geometry.

Future \cmb\ datasets offer the prospect of extending \Planck’s discovery reach. 
The \litebird\ satellite \cite{LiteBIRD:2020khw} is expected to deliver substantial improvements over \Planck\ for multipoles \(\ell \lesssim 200\), especially in polarisation. 
These gains should make scalar \(E\)-mode polarisation auto-correlation and cross-correlation with temperature more useful to searches for cosmic topology. 
Assessing the extent of this improvement and its implications for discovering or constraining cosmic topology is a target of current investigation by \compact\ in collaboration with \litebird. 
There appears to be, at least in principle, the potential to reach as far as \(d_{\mathrm{NC}} \approx 1.2 \, d_{\mathrm{LSS}}\) for all Euclidean manifolds and any \(S^{3}\) or \(H^{3}\) manifolds that are searched for.

Cosmic topology, by breaking statistical isotropy (and parity invariance), also alters the statistical character of the imprints of tensor fluctuations on \cmb\ temperature and polarisation. 
In the covering space, isotropy and parity enforce diagonality of all \(B\)-mode correlations in \(\ell\) and \(m\) and that \(B\)-mode polarisation is uncorrelated (except through the influence of foreground effects) with \(T\) and \(E\). 
In generic manifolds, \(T\), \(E\), and \(B\) are all correlated, and correlations are no longer diagonal in \(\ell\) and \(m\) \cite{COMPACT:2024cud, Samandar:2025kuf}. 
The consequences of this for detection of primordial tensor modes, and for the search for cosmic topology should primordial \(B\)-modes be discovered, is also a matter of ongoing investigation.

\section{Outlook: What Can We Hope to Measure?}
\noindent
Other present or future \cmb\ experiments may also be helpful for investigating cosmic topology. 
The \taurus\ super-pressure balloon experiment \citep{Taurus:2024dyi} aims to measure the \cmb\ \(E\)-mode angular power spectrum with high signal-to-noise ratio for multipoles \(\ell \lesssim 25\), and with up to \(70\%\) sky coverage. 
Scalar \(E\)-mode polarisation signals relevant to cosmic topology appear to be concentrated in precisely this multipole range \cite{COMPACT:2024cud}. 
\taurus\ may therefore approach \litebird\ in its discovery reach for cosmic topology. 
Past and current ground-based experiments, such as the Atacama Cosmology Telescope, the South Pole Telescope, and Simons Observatory, lack access to such low-\(\ell\) modes, but they do independently determine an enormous number of modes of temperature and polarisation at higher \(\ell\). 
No analysis has yet determined how this information is affected by cosmic topology, and thus neither how it might be used to probe cosmic topology nor how cosmic topology might affect the conclusions drawn for \flrw\ cosmology (for example, regarding tensor modes or non-Gaussianity).

Compared to the \cmb, far more
independent mode amplitudes are available if one can accurately measure the three-dimensional fields (e.g., the matter density and velocity, the metric and its time-derivative, etc.)\ that are the late-time ``descendants" of the fluctuations laid down in the early Universe. 
In consequence, there is, in principle, far more topologically relevant information available in sufficiently complete and high-precision galaxy surveys and line-intensity-mapping surveys that probe the entire observable volume of our Universe (see \cite{COMPACT:2022gbl}, and also Mihaylov, \textit{et al.} (COMPACT) 2026, in preparation). 
So far, it is not known how much of that information can be extracted with current or planned surveys, nor by how much that could improve our ability to probe cosmic topology. 

There is no guarantee that the topology of the Universe, whatever it may be, is discoverable. 
At least in our locale, the Universe is not small compared to the diameter of the last scattering surface. 
It may be extraordinarily large, in which case it seems unlikely that we will ever learn the topology of the Universe.
However, we may also be fortunate, and the evidence for cosmic topology may be sitting in existing data, or may be collected over the coming years in campaigns to observe the \cmb, to map the distribution and velocities of galaxies and other tracers of large scale structure, or to map the intensity of transition lines of atoms or molecules.
The monumental nature of a discovery of cosmic topology compels us to look as carefully as possible. 

\begin{acknowledgments}
\noindent
Several of us thank Jeffrey Weeks for valuable conversations over the years, 
as well as D.~Calvetti, E.~Somersalo, and D.~Singer. 
G.D.S.\ also thanks D.~Spergel, N.~Cornish, and J.R.~Bond for such conversations.

C.J.C., G.D.S., A.K., and D.P.M.\ acknowledge support from \nasa\ ATP grant 
RES240737 and from \nasa\ ADAP grant 24-ADAP24-0018. 
D.P.M.\ acknowledges support by the Bulgarian National Science Fund programme 
``VIHREN--2024'' project No. KP--06--DV/9/17.12.2024. 
A.N.\ is supported by the Richard S.\ Morrison Fellowship. 
A.S.\ and G.D.S.\ acknowledge support from DOE grant DESC0009946.

Y.A.\ acknowledges support by the Spanish Research Agency (Agencia Estatal de Investigaci\'on)'s grant RYC2020-030193-I/AEI/10.13039/501100011033, by the European Social Fund (Fondo Social Europeo) through the Ram\'{o}n y Cajal programme within the State Plan for Scientific and Technical Research and Innovation (Plan Estatal de Investigaci\'on Cient\'ifica y T\'ecnica y de Innovaci\'on) 2017-2020, by the Spanish Research Agency through the grant IFT Centro de Excelencia Severo Ochoa No CEX2020-001007-S funded by MCIN/AEI/10.13039/501100011033, by the Spanish National Research Council (CSIC) through the Talent Attraction grant 20225AT025, and by the Spanish Research Agency's Consolidaci\'on Investigadora 2024 grant CNS2024-154430.
J.C.D.\ is supported by the Spanish Research Agency (Agencia Estatal de Investigaci\'on), the Ministerio de Ciencia, Innovaci\'on y Universidades, and the European Social Funds through grant JDC2023-052152-I, as part of the Juan de la Cierva programme. 
M.M.B.\ acknowledges support by the Spanish Ministry of Science, Innovation and Universities under the FPU predoctoral grant FPU22/02306. 
F.C.G.\ is supported by Ministerio de Ciencia, Innovaci\'on y Universidades, Spain, through a Beatriz Galindo Junior grant BG23/00061 and by an UCOLIDERA grant from the Universidad de Córdoba.

A.H.J.\ acknowledges support from STFC in the UK\@. 
A.T.\ is supported by the European Union's Horizon Europe research and innovation 
programme under the Marie Sk\l{}odowska-Curie grant agreement No.\ 101126636. 
T.S.P.\ is supported by Funda\c{c}\~ao Arauc\'aria (NAPI Fen\^omenos Extremos do Universo, grant 347/2024 PD\&I).

We are thankful to the Istituto Nazionale di Fisica Nucleare in Italy. 
G.D.S.\ thanks the IFT (Madrid) and the University of Padua, and M.M.B.\ and C.P.\ thank CWRU for their hospitality.

This work made use of the High-Performance Computing Resource in the Core Facility for Advanced Research Computing at Case Western Reserve University and the facilities of the Ohio Supercomputing Center.
\end{acknowledgments}

\paragraph{Statement of Competing Interests} The authors declare no competing interests.

\paragraph{Author-contribution statement} 
D.P.M.\ and G.D.S.\ conceived the Review. 
G.D.S., D.P.M., and A.N.\ wrote the manuscript. 
C.J.C., D.P.M., A.N., and A.S.\ prepared the figures. 
All authors discussed and edited the manuscript.

\newcommand{\jcap}{J. Cosmol. Astropart. Phys.}

\end{document}